%% file: mgps-astroph.tex
\documentclass[useAMS,usenatbib]{mn2e}
\usepackage{graphicx}
\usepackage{rotating}
\usepackage{amsmath}
\usepackage{amssymb}

\usepackage{times}

\newcommand{\changes}[1]{{#1}}

\title[MGPS-2 Compact Source Catalogue]
      {The Molonglo Galactic Plane Survey (MGPS-2): Compact Source Catalogue}
\author[T. Murphy et al.]
       {T. Murphy$^{1,2}$\thanks{E-mail: tara@physics.usyd.edu.au},
        T. Mauch$^{1,3}$,
        A. Green$^{1}$\thanks{E-mail: agreen@physics.usyd.edu.au},
        R. W. Hunstead$^{1}$,
        B. Piestrzynska$^{1}$, \newauthor
        A. P. Kels$^{1,4}$, 
        P. Sztajer$^{1}$ \\
$^{1}$School of Physics, University of Sydney, NSW, Australia\\
$^{2}$School of Information Technologies, University of Sydney, NSW, Australia\\
$^{3}$Department of Physics, University of Oxford, UK\\
$^{4}$Department of Physics, University of Queensland, QLD, Australia
}

\date{Accepted 0000 December 07. Received 0000 December 07; in original form 0000 July 07}

\pagerange{\pageref{firstpage}--\pageref{lastpage}} 
\pubyear{2007}

\begin{document}

\maketitle

\label{firstpage}

\begin{abstract}
We present the first data release from the second epoch Molonglo Galactic 
Plane Survey (MGPS-2). MGPS-2 was carried out with the Molonglo 
Observatory Synthesis Telescope at a frequency of 843 MHz and with a restoring
beam of $45\arcsec\times45\arcsec\csc|\delta|$, making it the highest 
resolution large scale radio survey of the southern Galactic plane. 
It covers the range $|b| < 10\degr$ and $245\degr < l < 365\degr$ and 
is the Galactic counterpart to the Sydney University Molonglo Sky Survey 
(SUMSS) which covers the whole southern sky with $\delta \le -30\degr$ 
($|b| > 10\degr$).

In this paper we present the MGPS-2 compact source catalogue. 
The catalogue has 48\,850 sources above a limiting peak brightness of 
10 mJy beam$^{-1}$. Positions in the catalogue are accurate to 
$1\arcsec - 2\arcsec$. 
A full catalogue including extended sources is in preparation. 
We have carried out an analysis of the compact source density across the 
Galactic plane and find that the source density is not statistically higher
than the density expected from the extragalactic source density alone.

\changes{
We also present version 2.0 of the SUMSS image data and catalogue which 
are now available online.
The data consists of 629 $4.3^{\circ}\times 4.3^{\circ}$ mosaic images 
covering the $8100\,\rm{deg}^2$ of sky with $\delta \leq -30^\circ$ 
and $|b|>10^{\circ}$. The catalogue contains 210\,412 radio sources to a 
limiting peak brightness of 6 mJy beam$^{-1}$ at $\delta \leq -50^\circ$ 
and 10 mJy beam$^{-1}$ at $\delta > -50^\circ$. We describe the updates
and improvements made to the SUMSS cataloguing process. 
}
\end{abstract}


\section{Introduction}
The Galactic plane exhibits radio emission on a wide range of angular scales.
To study the range of structures seen, from ultra-compact sources to complex 
diffuse sources, a survey with both high resolution and good sensitivity to
emission on large angular scales is essential.
The second epoch Molonglo Galactic Plane Survey (MGPS-2) has better resolution 
(restoring beam of $45\arcsec\times45\arcsec\csc|\delta|$) and sensitivity than 
any previous large scale radio surveys of the southern Galaxy. 
It is the Galactic counterpart to the Sydney University Molonglo Sky Survey 
\citep[SUMSS;][]{bock99,mauch03} and between them they cover the whole sky 
south of $\delta = -30\degr$ at a frequency of 843 MHz.
MGPS-2 was carried out over the same period of time as SUMSS, also using the 
Molonglo Observatory Synthesis Telescope \citep[MOST;][]{mills81,robertson91}.
This paper presents the first data release and compact source catalogue from 
MGPS-2. 

The precursor to MGPS-2 was the original Molonglo Galactic Plane Survey
\citep[MGPS;][]{green99a} which was carried out at 843 MHz over an eleven 
year period from 1983 to 1994. This covered the region 
($|b| < 1.5\degr$, $245\degr < l < 355\degr$), at a resolution of 
$43\arcsec\times43\arcsec\csc|\delta|$.
After a major upgrade to extend the diameter of the MOST field of view 
from $70\arcmin$ to $163\arcmin$, MGPS-2 was commenced in 1997. 
It extends the coverage of the MGPS in Galactic latitude to $|b| < 10\degr$ 
(with $245\degr < l < 365\degr$, equivalent to $\delta = -30\degr$).
The boundary between MGPS-2 and SUMSS of $|b| = 10\degr$ was chosen 
so that the coverage of MGPS-2 is roughly equal to the Zone of Avoidance in 
which it is difficult to study extragalactic sources due to extinction and 
confusion from the Galaxy. 
At the time the survey was designed, this was defined as approximately 
$|b| \le 10-12\degr$ (see, for example, \citealt{lu90}).
Throughout this paper we will refer to the original Molonglo Galactic 
Plane Survey as MGPS-1, with the new survey being discussed as MGPS-2.
An overview of the MGPS-2 survey design is given in \citet{green99b}.
Table \ref{t_compare} compares the properties of MGPS-1 and MGPS-2.

Although the extragalactic radio sky is well covered by several surveys at mJy 
levels --- NVSS \citep{condon98}, SUMSS \citep{mauch03} and 
FIRST \citep{becker95} --- there have been comparatively few radio surveys of 
the Galactic plane at equivalent resolution, sensitivity and uv-coverage.
The International Galactic Plane Survey has imaged HI and 21 cm continuum 
emission across the entire Galactic plane. It consists of three individual
surveys: the Canadian Galactic Plane Survey \citep[CGPS;][]{taylor03}, the 
VLA Galactic Plane Survey \citep[VGPS;][]{stil06} and the Southern Galactic 
Plane Survey \citep[SGPS;][]{mcclure-griffiths05}. These surveys combine 
single-dish and interferometer data to achieve high (arcminute) resolution 
while probing a wide range of spatial scales.
The Multi-Array Galactic Plane Imaging Survey \citep[MAGPIS;][]{helfand06} 
is the highest resolution large scale radio survey of the Galactic plane 
to date. It has an angular resolution of $6\arcsec$ and covers the region 
($5\degr < l < 32\degr$, $|b| < 0\fdg8$ in the first Galactic quadrant.
Table \ref{t_surveys} compares MGPS-2 with other radio surveys of the 
Galactic plane.
\begin{table}
\centering
\caption{A comparison of the MGPS-1 and MGPS-2 surveys\label{t_compare}}
\begin{tabular}{lll}\hline
  & MGPS-1 & MGPS-2 \\
\hline
Longitude & $245\degr < l < 355\degr$ & $245\degr < l < 365\degr$ \\
Latitude & $|b| \le 1.5\degr$ & $|b| \le 10\degr$ \\
Area surveyed & $330\degr^2$ & $2400\degr^2$ \\
Field size & $70\arcmin\times70\arcmin\csc|\delta|$ & $163\arcmin\times163\arcmin\csc|\delta|$\\
Number of fields & 455 & 621 \\
\hline
\end{tabular}
\end{table}

\begin{table*}
\centering
\caption{Comparison of Galactic plane surveys at radio wavelengths.\label{t_surveys}}
\begin{tabular}{llllll}\hline
Survey & Frequency & Longitude & Latitude & Resolution & Reference \\
\hline
7C(G) & 151 MHz & $80\degr < l < 104\degr$ & $|b| \le 5\degr$ & $70\arcsec\times70\arcsec\csc|\delta|$ & \cite{vessey98} \\
 &  & $126\degr < l < 180\degr$ & & & \\
CGPS & 408 MHz & $74\fdg2 < l < 147\fdg3$ & $-6\fdg7 < b < +8\fdg7$ & $3\farcm4\times3\farcm4\csc\delta$ & \cite{taylor03} \\
MGPS-1 & 843 MHz & $245\degr < l < 355\degr$ & $|b| < 1.5\degr$ & $43\arcsec\times43\arcsec\csc|\delta|$ & \cite{green99a} \\
MGPS-2 & 843 MHz & $245\degr < l < 365\degr$ & $|b| < 10\degr$ & $43\arcsec\times43\arcsec\csc|\delta|$ & This paper \\
CGPS & 1420 MHz & $74\fdg2 < l < 147\fdg3$ & $-3\fdg6 < b < +5\fdg6$ & $1\arcmin\times1\arcmin\csc\delta$ & \cite{taylor03} \\
SGPS I  & 1420 MHz & $253\degr \le l \le 358\degr$ & $|b| < 1\fdg5$ & $\sim 2\arcmin$ & \cite{haverkorn06} \\
SGPS II & 1420 MHz & $5\degr \le l \le 20\degr$ & $|b| < 1\fdg5$ & $\sim 2\arcmin$ & \cite{mcclure-griffiths05} \\
VGPS & 1420 MHz & $18\degr < l < 67\degr$ & $|b|<1\fdg3$ to $|b|<2\fdg3$ & $1\arcmin$ & \cite{stil06} \\
Effelsberg & 2695 MHz & $358\degr \le l \le 76\degr$ & $|b| \le 5\degr$ & $9\farcm4$ & \cite{reich90a} \\
 & & $76\degr \le l \le 240\degr$ & & & \cite{furst90} \\
MAGPIS & 1400 MHz & $5\degr < l < 32\degr$ & $|b| < 0\fdg8$ & $6\arcsec$ & \cite{helfand06} \\
\hline
\end{tabular}
\end{table*}

\changes{
Some of the major science goals of MGPS-2 are: to search for young 
supernova remnants and supernova remnants away from the Galactic plane, 
to search for pulsar wind nebulae and pulsar counterparts and to use the 
two-epoch data to study long term variability. Observations of 
individual MGPS-1 and MGPS-2 fields are separated by 3$-$20 years, which gives
us a range of variability timescales. In addition we have made repeated 
observations of several fields on shorter timescales of weeks to months. 
}

One aim of producing a compact source catalogue is to investigate the source 
density across the plane. \citet{whiteoak92a} indicated 
there may be a small excess of non-thermal sources. We investigate this 
further in Section \ref{s_density}. Another aim is that we can detect
extragalactic sources in  the Zone of Avoidance, that would be undetectable
at shorter wavelengths.
In addition, the MGPS-2 compact source catalogue complements the SUMSS catalogue
released by \citet{mauch03}.

In Section \ref{s_obs} we give an overview of the survey observations and 
data reduction. Section \ref{s_cat} describes the algorithms used for 
source detection, and the construction of the compact source catalogue. 
We include a description of the SUMSS catalogue as the surveys are closely 
linked. In Section \ref{s_acc} we present an analysis of the catalogue 
accuracy and reliability and Section \ref{s_format} gives the catalogue format.
Finally, Section \ref{s_analysis} contains our analysis of the data, 
including the compact source density distribution.

\section{Observations}\label{s_obs}
The MOST is an east-west array and so requires 12 hr for a full synthesis. 
The field of view and synthesised beam are elliptical, being elongated by a 
factor of $\csc|\delta|$ in Declination. 
The detailed technical specifications of the MOST are summarised in 
Table \ref{t_most}.
\begin{table}
\centering
\caption{Technical specifications of the MOST. Note that the Declination range
given is for a fully synthesised image.\label{t_most}}
\begin{tabular}{lr}\hline
Parameter & Value \\
\hline
Centre frequency & 843 MHz \\
Bandwidth & 3 MHz \\
Declination range & $-30\degr$ to $-90\degr$\\
Restoring beam & $45\arcsec\times45\arcsec\csc|\delta|$ \\
Field size & $163\arcmin\times163\arcmin\csc|\delta|$ \\
Dynamic range (processed data)$^*$ & 250:1 \\
System noise $1\sigma$ (after 12 hr) & 1$-$2 mJy beam$^{-1}$ \\
Polarisation & RHC \\
\hline
\end{tabular}
\medskip

$^*$ The dynamic range in regions away from bright sources.\vspace{3mm}
\end{table}

\begin{figure}
\centering
\includegraphics[width=8cm]{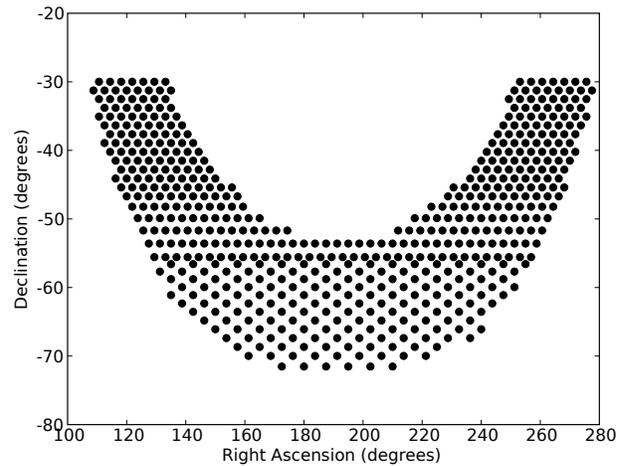}
\caption{The hexagonal pointing centre grid for MGPS-2. This was chosen to
optimise speed of coverage.\label{f_point}}
\end{figure}

The pointing pattern used for MGPS-2 is shown in Figure \ref{f_point}. 
The pointing centre grid is different to the SUMSS survey; in SUMSS, the 
pointing centres were chosen to give uniform signal to noise across all 
mosaics. However, in the Galactic plane, the limiting factor in the data 
quality comes from the complexity of Galactic sources, the dynamic range of 
the telescope, and the limitations in the data reduction for complex sources. 
\changes{
Because of this, we make speed of coverage a priority rather than uniform
signal to noise and so there is minimal overlap between the fields in the 
hexagonal pointing grid of MGPS-2.
}
As a result the signal to noise is not as uniform across MGPS-2 as it is 
for SUMSS.

The rms noise in individual MOST images is a combination of thermal noise 
and source confusion from both the main beam and sidelobes of the MOST. 
\changes{
The noise level varies with Declination, position in the field and 
operational status of the telescope. 
The precise operational condition of the telescope electronics and feeds 
may change during a 12-hour run. Since the telescope is calibrated only at 
the beginning and end of each run, there may be changes which have a 
small effect on the sensitivity of any given observation.
}
In the centre of a MOST image the typical rms noise is $\sim 1$ mJy 
beam$^{-1}$ for quiet fields away from bright complex regions. 
However, in MGPS-2 the sensitivity is dynamic range limited rather than
noise limited.

\changes{
The MOST data are reduced using a custom process described in \citet{bock99}. 
The calibration of the images in both position and flux density is based on
observations of a sample of strong unresolved sources (taken from 
\citet{campbell-wilson94}) that are made before and after each 12 hour 
synthesis observation.
}

\begin{figure*}
\begin{center}
\includegraphics[width=14cm,angle=270]{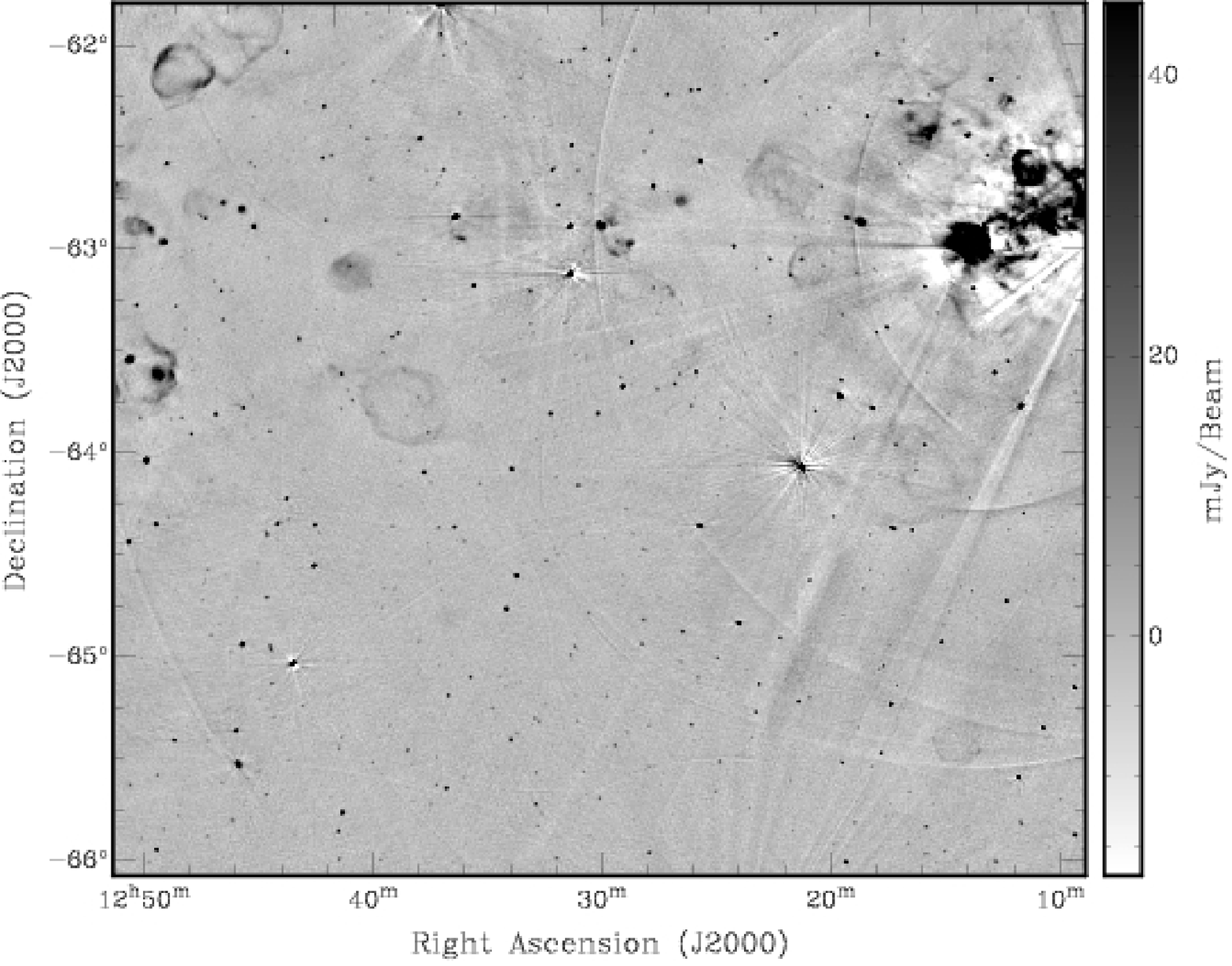}
\caption{A typical MGPS-2 mosaic J1230M64, centred at J2000 
$(\alpha, \delta) = 12^h30^m, -64\degr$, approximately equivalent 
to $(l, b) = 279\degr20\arcmin, -01\degr41\arcmin$. 
The full MGPS-2 survey consists of 130 of these $4\fdg3\times4\fdg3$ mosaics.
\label{f_mosaic}}
\end{center}
\end{figure*}

The individual images are combined to make $4\fdg3\times4\fdg3$ mosaics. 
This was done to show large scale source distribution and to match with SUMSS. 
A typical MGPS-2 mosaic J1230M64\footnote{The naming scheme for MGPS-2 mosaics
is J{\it hhmm}M{\it dd} where J signifies J2000 coordinates, {\it hhmm} 
is the RA in hours and minutes of the mosaic centre, 
M signifies southern Declination and {\it dd} is the absolute value of the 
Declination of the mosaic centre in degrees. 
This has been kept in equatorial coordinates for consistency with SUMSS.}, 
is shown in Figure \ref{f_mosaic}.
Images covering the central $4\degr$ strip of MGPS-2 ($|b| < 2\degr$) 
are shown in Figures \ref{f_strips1} and \ref{f_strips2}. 
The pixel size in the final mosaics is $11\arcsec\times11\arcsec\csc|\delta|$.
Note that while the synthesised beam of the MOST is $43\arcsec\times43\arcsec\csc|\delta|$, 
the MGPS-2 images are restored with a resolution of $45\arcsec\times45\arcsec\csc|\delta|$.

\section{SUMSS}
The basic cataloguing method used for MGPS-2 is the same as for the SUMSS 
catalogue. This was done to ensure as much uniformity between the 
catalogues as possible. The process is explained fully in \citet{mauch03} 
but we summarise it here. 
We also give an update on the status of the SUMSS catalogue, which has been
updated and improved since \citet{mauch03}. 
This is discussed in Section \ref{s_status}.

\subsection{The SUMSS catalogue}\label{s_sumss}
The outputs from the custom MOST data reduction process are 
$4\fdg3\times4\fdg3$ mosaics. The SUMSS cataloguing software runs 
the {\sc aips} tool {\sc vsad} (written for the NVSS survey; \citealt{condon98})
over these mosaics to produce a preliminary source list.
{\sc vsad} fits an elliptical Gaussian to all points above a given cutoff. 
The parameters of each Gaussian returned by {\sc vsad} are the J2000 right 
Ascension $\alpha$ and Declination $\delta$ (degrees), peak brightness 
$A_{843}$ (mJy beam$^{-1}$), total flux density $S_{843}$ (mJy), 
FWHM fitted source major and minor axes $\theta_M,\theta_m$ (arcseconds) 
and the fitted position angle of the major axis (degrees east from north).

\begin{figure*}
\includegraphics[angle=270,width=5.7cm]{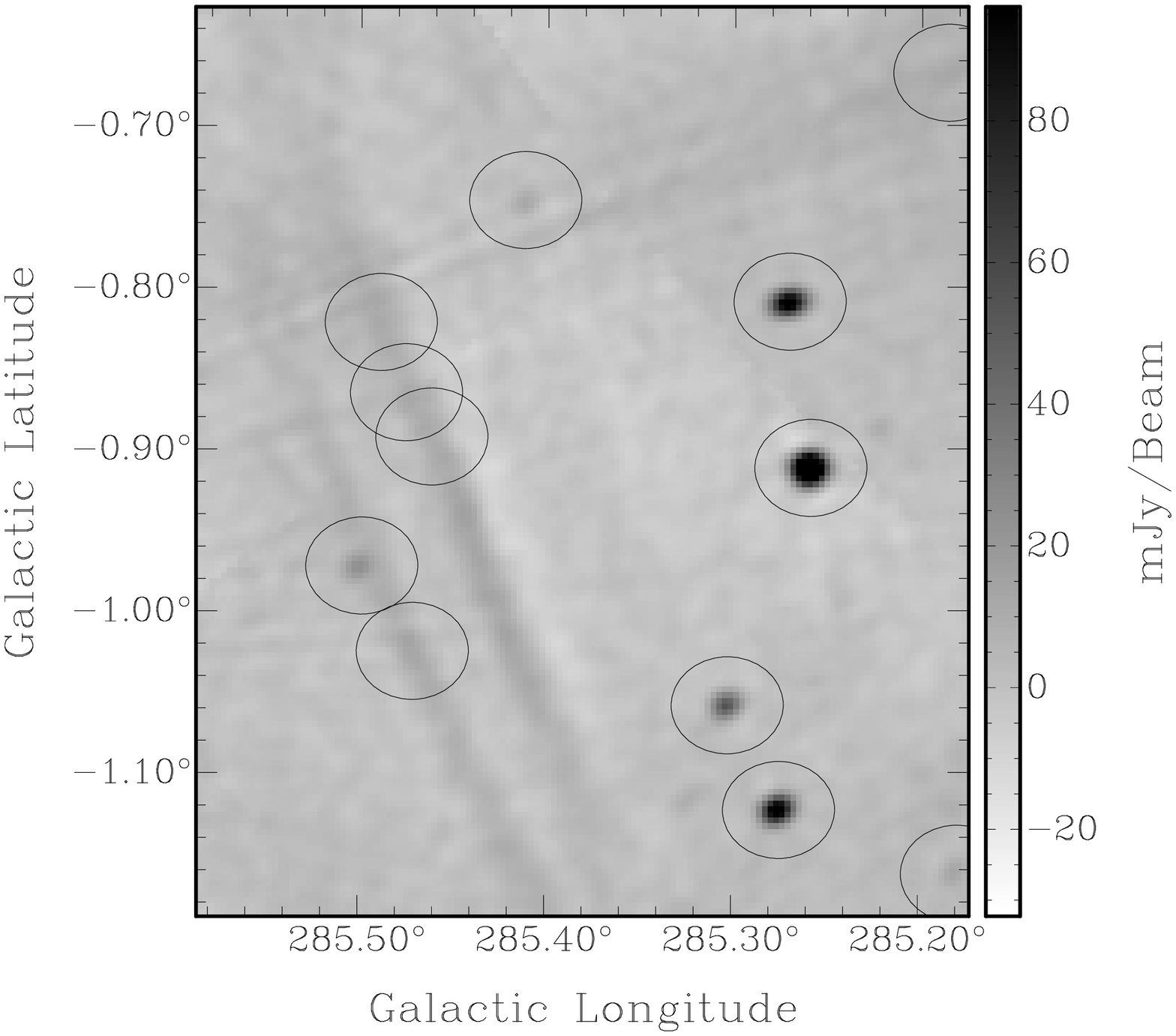}
\includegraphics[angle=270,width=5.7cm]{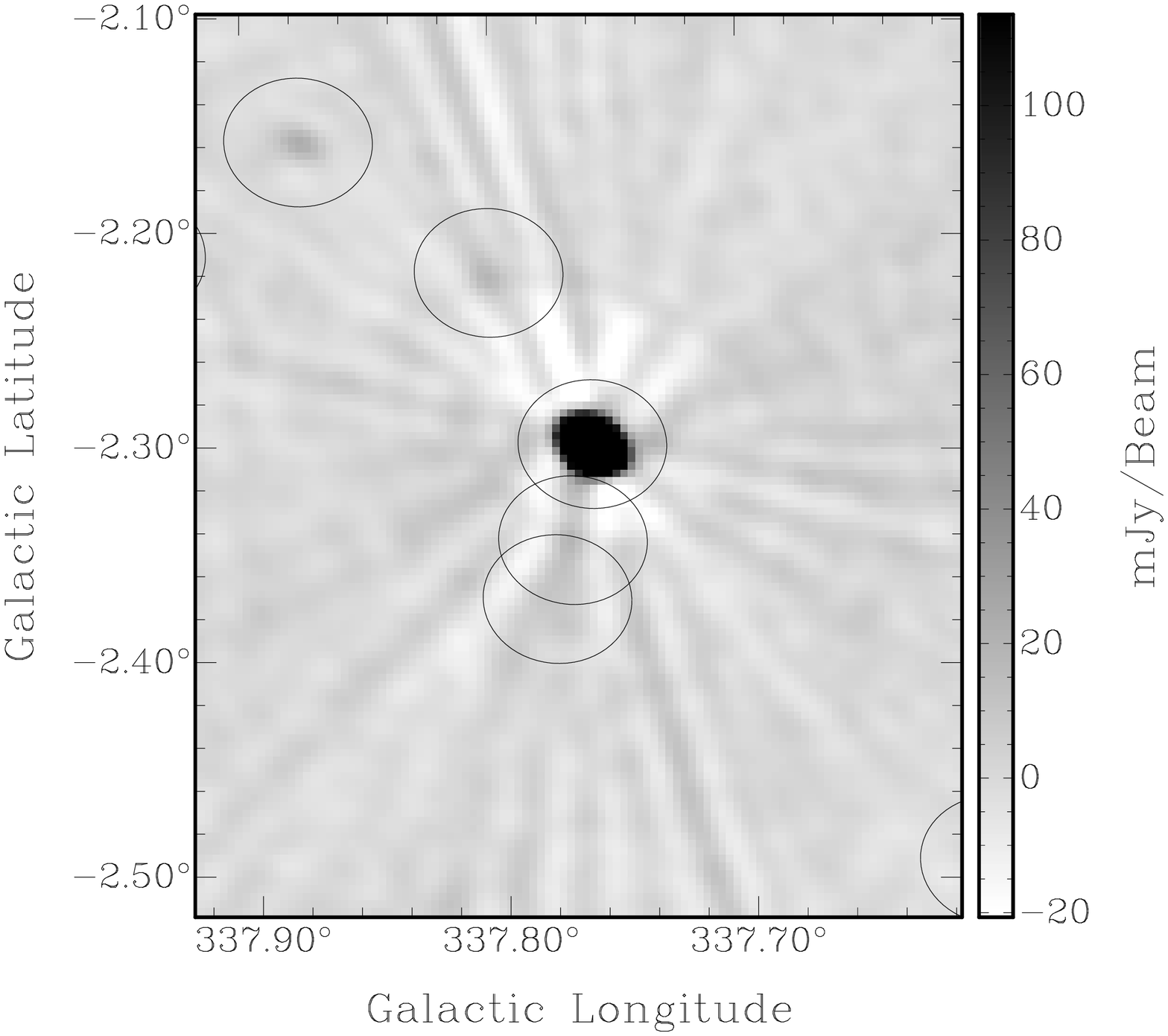}
\includegraphics[angle=270,width=5.7cm]{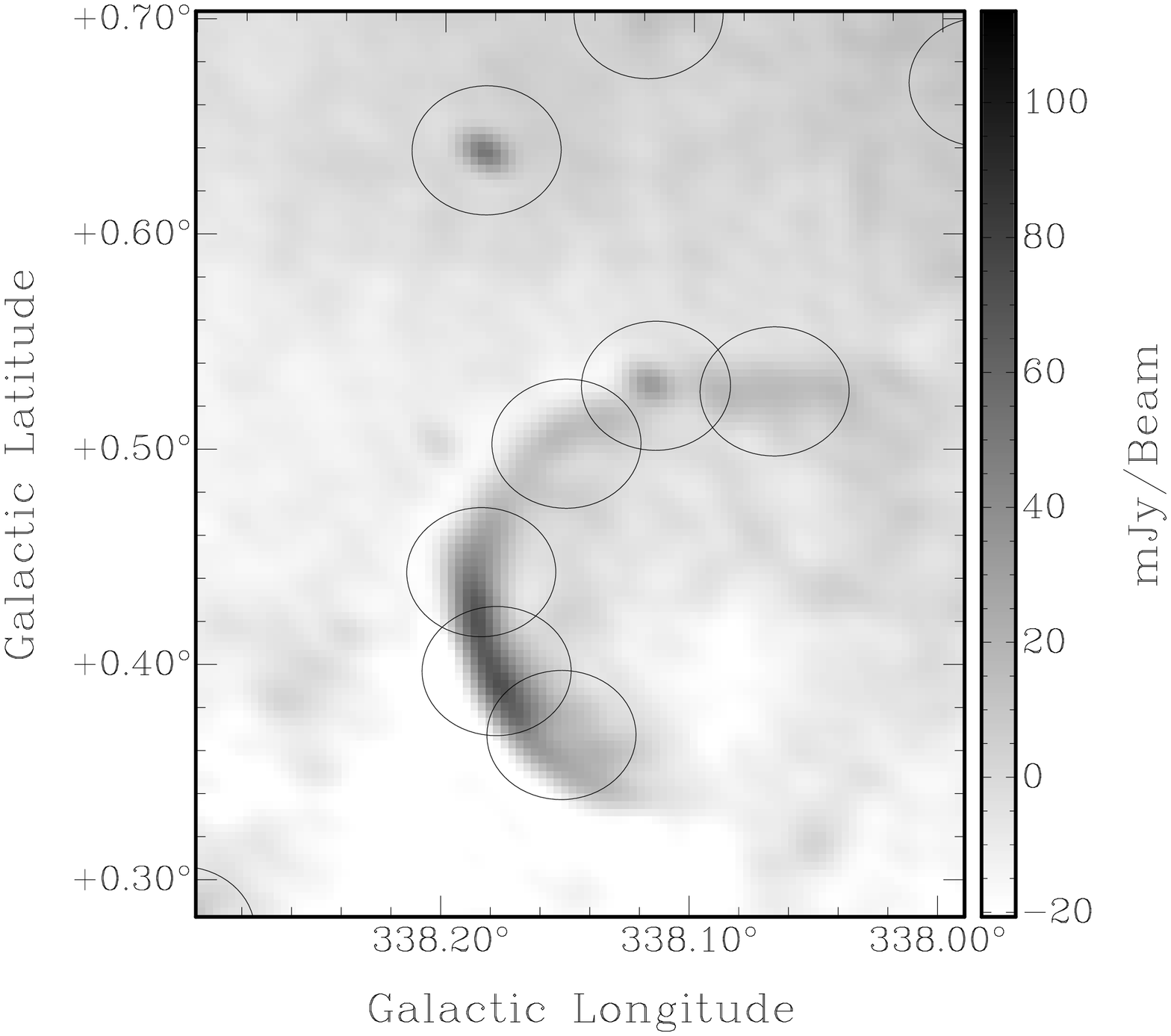}
\caption{Three images showing the variety of sources and artefacts present 
in MGPS-2 mosaics which are fitted as spurious sources by {\sc vsad}. 
In each case, candidate sources fitted by {\sc vsad} are shown by the
annotation ellipses. The left image shows part of a grating ring associated 
with a nearby strong source. The middle image shows a source with
an integrated flux of 1.4 Jy and 
the radial spikes associated with it. The spikes here have peak amplitude 
of $\sim 20$ mJy beam$^{-1}$ and extend about $30\arcmin$ away 
from the source. The right image shows a diffuse extended source in the 
mosaic J1640M48 which is fitted as multiple sources by {\sc vsad}.}
\label{f_artefacts}
\end{figure*}

As discussed in Section 3 of \cite{mauch03} the SUMSS and MGPS-2 mosaics 
have various artefacts in them. These are caused by {\it grating rings} 
and {\it radial spokes} and are a problem for cataloguing because {\sc vsad} 
incorrectly extracts multiple point sources for these artefacts.
Examples of these artefacts are shown in Figure \ref{f_artefacts}.

\citet{mauch03} presented the solution to this problem, which was to train 
a decision tree classifier on a set of manually classified sources, and 
then apply it to the rest of the catalogue. They estimated the classification 
accuracy of the decision tree at around 96 per cent. More details of the 
cataloguing process are discussed in \cite{mauch03}, including the 
process for removing duplicate sources from the regions in which the 
mosaics overlap.

\subsection{SUMSS status}\label{s_status}
SUMSS observations were completed in early 2007 and a full complement
of 629 $4.3^{\circ}\times 4.3^{\circ}$ mosaic images covering the 
$8100\,\rm{deg}^2$ of sky with $\delta \leq -30^\circ$ and $|b|>10^{\circ}$ 
have been made available online. A full release of the SUMSS catalogue was 
built on 2007 August 8 using all of the released mosaics.  
It contains 210\,412 radio sources to a limiting peak 
brightness of 6 mJy beam$^{-1}$ at $\delta \leq -50^\circ$ and 
10 mJy beam$^{-1}$ at $\delta > -50^\circ$.
We have made minor improvements and changes to the catalogue pipeline since
the previous version described in \citet{mauch03}. These changes are as 
follows:

\begin{enumerate}
\item As reported by \citet{klamer06}, around 20 per cent 
of the flux densities in the region $\delta \geq -50^\circ$
are larger than expected by a few per cent. This problem was related to 
the ellipticity of the MOST beam at more northern declinations and arises
from an assumption in {\sc vsad} which was optimised for the circular
beam of the NVSS survey. We have corrected this problem and revised the
flux densities of affected sources in the catalogue.

\item We have revised the decision tree and added a further 1000 sources 
to the training set. The original decision tree was found to 
be performing poorly for rare source types with no examples
in the training set. We have visually inspected SUMSS images extensively
and believe the decision tree now performs well across the entire survey. The
results of decision tree classification
listed in Table 1 of \citet{mauch03} remain the same and we estimate
its accuracy to be 96 per cent.

\item We have added one further large
diffuse source (IC\,4296) which was fitted badly by {\sc vsad}. 
As with the sources added by hand in earlier versions of the SUMSS 
catalogue we have not assigned a peak amplitude or a source size, 
only a position and total flux density. The flux densities of these sources
have been calculated by summing the the pixels inside a manually defined
source area.
\end{enumerate}

The present version of the SUMSS catalogue (version 2.0) supersedes all 
previous versions and should be used whenever possible.
Users of prior catalogue releases should note that the described 
changes to the pipeline may have caused some measured positions or 
flux densities to change by more than their previously quoted error.

\section{Catalogue construction}\label{s_cat}
Extracting sources from astronomical images is a well studied problem.
Most of the existing tools (e.g. {\sc vsad}, {\sc imsad}, SeXtractor) 
work well for compact sources, but are not adequate for more diffuse
and extended sources. Hence the catalogue creation process is more 
involved when dealing with images of the Galactic plane, which is dominated 
by extended, diffuse objects.

In the previous section we described the SUMSS cataloguing process used in 
\cite{mauch03} which is the basis for our MGPS-2 cataloguing process. 
Here we describe the modifications to the method required for the MGPS. 

\subsection{The MGPS-2 catalogue}\label{s_mgpscat}
In the SUMSS catalogue about 90 per cent of sources were unresolved point 
sources. Most of the remaining sources were only slightly extended,
with only 1 per cent being resolved along both axes. 
\changes{Hence the cataloguing process described above, based largely on the 
point source extraction tool {\sc vsad}, performed well.
For the region $2\degr < |b| < 10\degr$ the MGPS-2 compact source catalogue 
was produced  using the unmodified SUMSS cataloguing software, in which
the accuracy of the source classification is estimated to be around 
96 per cent \citep{mauch03}.}

\begin{figure}
\begin{center}
\includegraphics[angle=270,width=\linewidth]{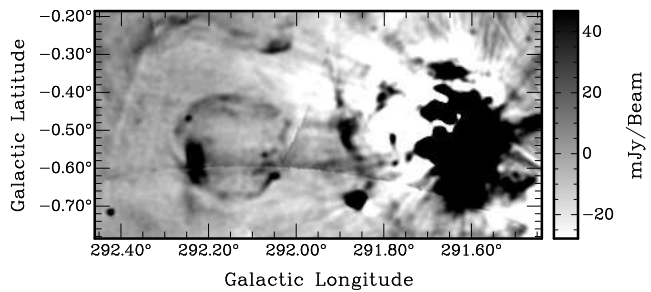}
\caption{Image showing the negative bowls that occur around 
bright extended sources. The bright source to the right of the image 
has a peak flux density of $\sim6$ mJy beam$^{-1}$, and an integrated
flux of $\sim150$ Jy. The negative bowl
extends to $\sim-0.2$ mJy beam$^{-1}$.
\label{f_trough}}
\end{center}
\end{figure}

However, within 2 degrees of the Galactic plane there are a large number of 
extended and diffuse sources which makes the cataloguing for MGPS-2 less 
straightforward.
There are also a large number of negative bowls around bright extended 
sources, which are an observational artefact. As explained in 
\cite{green99a} the MOST does not measure the autocorrelation of its elements,
or correlations on baselines shorter than $42.9\lambda$. The effect of this
is that structures on angular scales larger than $\sim20\arcmin-30\arcmin$ 
are not detected and all sources are surrounded by a `bowl' of negative 
flux density.
The CLEANing process takes care of this for all but the brightest sources,
by interpolating for the unmeasured short spacings. Hence in the final images, the 
negative bowls remain around sources with an integrated flux density of 
$\gtrsim 100$ mJy beam$^{-1}$. This is illustrated in 
Figure \ref{f_trough}.

\begin{figure}
\begin{center}
\includegraphics[angle=270,width=\linewidth]{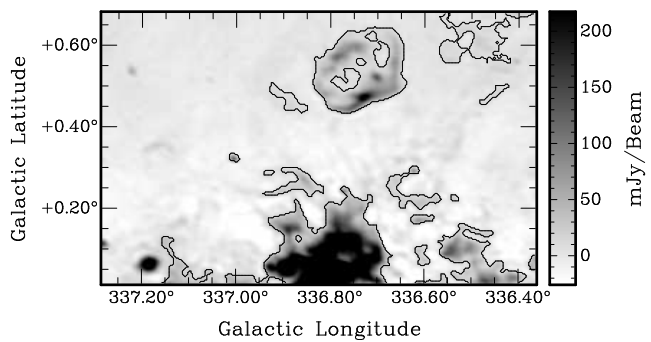}
\caption{Results from running the flood fill algorithm. The image shows
a small region of mosaic J1640M48, containing several large diffuse sources. 
The contours show the boundary of the {\it filled} region in which all sources
found by {\sc vsad} will be excluded from the final catalogue.\label{f_flood}}
\end{center}
\end{figure}

Extended sources present additional problems as {\sc vsad} identifies multiple 
point sources which are actually part of the same extended source. 
To solve this we developed new cataloguing software based on the flood fill 
algorithm, which works as follows:
\begin{enumerate}
\item Find each pixel above the peak cutoff level.
\item Explore the region around each peak pixel, `filling' the pixels 
around the peak in each direction until a minimum cutoff level is reached.
\item Each filled region is then considered a source.
\end{enumerate}
\changes{Note that the peak cutoff level used is constant across the survey 
and is not adjusted for local variations in base level.}
An example of the results is shown in Figure \ref{f_flood}.
This process extracts all the sources independently from {\sc vsad} and can 
then be used in combination with the {\sc vsad} identification.
Each source from the preliminary catalogue (produced with the SUMSS 
cataloguing software) is associated with a filled region. 
If the size of that region is significantly larger than a beam size, the 
source is classified as extended. 
Otherwise it is classified as compact.
For the purpose of the present source catalogue, only the latter sources
are included.
Extended sources will be dealt with in a subsequent publication. 
In extremely complex regions we will use visual inspection and 
cross-matching with source catalogues at other frequencies 
(e.g. mid-infrared) to identify the sources.

For the region $|b| < 2\degr$ the MGPS-2 compact source catalogue was produced 
using the method outlined above. We checked a randomly selected subset of 
sources by hand and from that we estimate the accuracy of this method to be 
about 93 per cent. This is quite variable depending on the proximity 
of Galactic plane emission. In regions away from the complex sources the accuracy is 
typically around 96 per cent whereas in complex Galactic plane regions it can 
be as low as 86 per cent. 

In addition, there are three regions of significant diffuse emission that 
occur outside the $|b| < 2\degr$ region. These are caused by the Puppis-A 
supernova remnant, the Vela supernova remnant and the radio galaxy PKS 
1610$-$60. In these regions we manually classified the sources for the 
compact source catalogue. 

\subsection{Noise}\label{s_noise}
The rms noise varies throughout the MGPS-2 mosaics. In particular there is 
higher background noise in regions around bright sources. Away from the 
plane this is quite localised, but close to the plane ($|b|<2\degr$) this 
extends throughout most of the mosaics due to the presence of a large 
number of bright, extended sources.

To determine a threshold below which sources are discarded from the 
catalogue, we estimated the median rms noise over the whole survey 
region. We calculated the rms noise around each source by defining 
a $25\times25$ pixel box centred on the source, and fitting a histogram 
to the pixels in the residual image. The residual image is produced by 
{\sc vsad} during the cataloguing process and consists of the original
mosaics with all detected sources subtracted. 

The median rms noise in the region $|b| > 2\degr$ is 
1.6 mJy beam$^{-1}$. In the region $|b| \le 2\degr$ it is 
2.1 mJy beam$^{-1}$. Using a $5\sigma$ 
limit for the catalogue gives us a catalogue flux density threshold of 
10 mJy beam$^{-1}$. We have chosen this relatively conservative 
limit of {\it at least} $5\sigma$ because the rms noise in the 
images can increase significantly close to bright extended sources.

A small number of sources have an rms noise significantly higher than the
median values quoted about. These are mostly due to a neighbouring bright 
source with artefacts (as discussed in Section \ref{s_sumss}).

\section{Accuracy}\label{s_acc}
The errors given in the catalogue are a combination of fitting 
uncertainties and calibration uncertainties of the MOST.
The fitting uncertainties were determined in the same way as for the SUMSS and 
NVSS catalogues, using equations derived in \cite{condon97}.
\changes{
For weaker sources ($S_{843} < 50$ mJy) the calibration uncertainties of 
the MOST are negligible and the fitting uncertainties dominate.
For stronger sources ($S_{843} \ge 50$ mJy) the calibration uncertainties
dominate.
}
The calculations for MGPS-2 follow those of SUMSS. 
The process is discussed in detail in Section 4 of \citet{mauch03} but 
it is outlined again here for completeness.

\subsection{Positional uncertainties}
The fitting variances in the source positions are given by:
\begin{equation}
\sigma{_\alpha^2}=\sigma{_{\rm{M}}^2}\sin^2\left({\rm
    PA}_{\rm F}\right)+\sigma{_{\rm{m}}^2}\cos^2\left({\rm PA}_{\rm F}\right),
\label{raerr}
\end{equation}
\begin{equation}
\sigma{_\delta^2}=\sigma{_{\rm{M}}^2}\cos^2\left({\rm
    PA}_{\rm F}\right)+\sigma{_{\rm{m}}^2}\sin^2\left({\rm PA}_{\rm F}\right),
\label{decerr}
\end{equation}
where the rms noise-like uncertainties of the fitted major and minor axes
($\sigma{_{\rm{M}}}$ and $\sigma{_{\rm{m}}}$, respectively) are derived using 
equation~25 of \citet{condon97}. PA$_{\rm F}$ is the fitted position angle 
of the source in degrees east of north.

\begin{figure}
\begin{center}
\includegraphics[width=\linewidth]{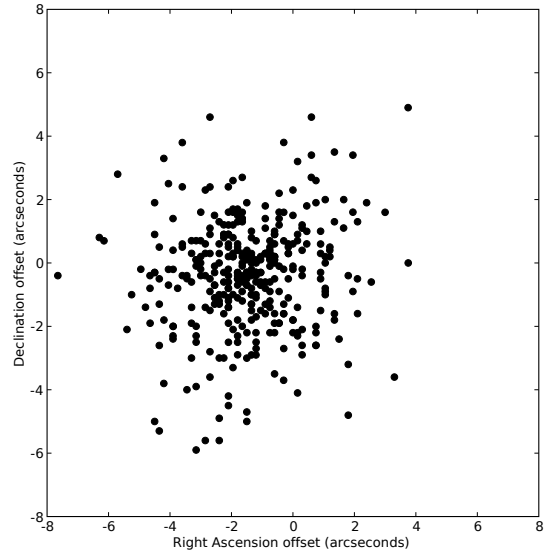}
\caption{Positional offsets in Right Ascension and 
Declination (MGPS-2 position minus NVSS position)
for 356 strong point sources in the overlap region between these 
surveys.\label{f_offset}}
\end{center}
\end{figure}

We have determined the calibration uncertainty 
for strong sources ($S_{843} > 200$ mJy) by repeating the analysis of 
\citet{mauch03} for the MGPS-2 region. This was done by comparing the 
MGPS-2 positions with those from the NVSS catalogue in the overlap region 
between MGPS-2 and NVSS ($-40\degr < \delta < -30\degr$). The positions
of strong sources in the NVSS catalogue are accurate to within 
$(\epsilon_\alpha,\epsilon_\delta) = (0\farcs45,0\farcs56)$ \citep{condon98}. 
For this analysis we used only point sources to avoid the larger positional 
errors associated with extended sources.
We define a point source as one in which the fitted major ($\theta_M$) and 
minor ($\theta_m$) axes of the source do not exceed 
$2.33\sigma(\theta_M,\theta_m)$. This definition is explained in the next 
section. There are 356 sources that meet these criteria, all of which had 
an NVSS counterpart. Figure \ref{f_offset} shows the offsets in Right 
Ascension ($\Delta \alpha$) and Declination ($\Delta \delta$) between NVSS 
and MGPS-2. The mean offsets are $\langle\Delta \alpha\rangle = -1\farcs1$ 
and $\langle\Delta \delta\rangle = -0\farcs3$.

The mean RA offset of $-1\farcs1$ is of some concern, and different to the 
value reported in \citet{mauch03}. We repeated the analysis with bright 
SUMSS point sources in the SUMSS$-$NVSS overlap region and found mean offsets 
of $\langle\Delta \alpha\rangle = -1\farcs0$ and 
$\langle\Delta \delta\rangle = -0\farcs5$.
As an additional check we compared the positions of the International 
Celestial Reference Frame (ICRF) defining calibrators \citep{fey04}
(of which there are 40 with $\delta < -30\degr$) with the corresponding 
SUMSS and MGPS-2 sources. This gave mean offsets of 
$\langle\Delta \alpha\rangle = -0\farcs6$ and 
$\langle\Delta \delta\rangle = -0\farcs1$.
However, the RA offset results showed a trend with Declination, with 
$\langle\Delta \alpha\rangle = -0\farcs2$ for $\delta < -50\degr$ and 
$\langle\Delta \alpha\rangle = -1\farcs1$ for $\delta > -50\degr$. 
This suggests that the underlying problem is Declination dependent, and 
the positional accuracy is worst in the NVSS overlap region. 

The rms of the offsets between the MGPS-2 and ICRF catalogues have been
used as a measure of the MGPS-2 positional uncertainties.
These are $\epsilon_\alpha = 0\farcs9$ for Right Ascension and 
$\epsilon_\delta = 0\farcs9$ for Declination.
However, there is also a systematic positional offset which depends on 
Declination and can be as high as 
$(\Delta\alpha,\Delta\delta) = (-1\farcs0,-0\farcs5)$.
This problem is likely to be caused by something in the telescope itself or 
the reduction software and could not be isolated by our initial investigation.
The problem is consistent with a small error in the time stamp of each sample,
of order 0.1 seconds.
\changes{As we do not fully understand the cause of this offset, and to remain
consistent with the SUMSS catalogue, we have decided to not to remove
this offset from the published positions.}

\subsection{Source sizes}
To determine if a source in the catalogue is significantly resolved we
compare the errors ($\sigma(\theta_M)$, $\sigma(\theta_m)$) of the fitted 
major and minor axes ($\theta_M$, $\theta_m$) for each source with the 
size of the beam. As discussed in \citet{condon98}, the probability that 
the fitted size of a point source would be larger than the beam by
more than $2.33\sigma(\theta_M,\theta_m)$ is less than 2 per 
cent. Hence we compare the beam plus $2.33\sigma(\theta_M,\theta_m)$ to 
determine if a source is resolved along either axis.

Sources for which either axis is believed to be resolved are then deconvolved 
and values for the deconvolved major axis, minor axis and position angles 
are given in the catalogue. No source sizes are given for unresolved sources. 
Users of the catalogue should note that the deconvolved source sizes are
only indicative of whether the source is extended or not. Checking the 
original images is recommended for confirmation of detailed source properties.

\subsection{Flux densities}\label{s_flux}
There are two main sources of error in the quoted flux densities: the
relative flux density calibration uncertainty of the MOST 
($\epsilon_{A_{843}}$) and the local noise uncertainty ($\sigma_{A_{843}}$).
The local noise uncertainty is calculated as
\begin{equation}
\sigma^2_{A_{843}} = \frac{A^2_{843}}{\rho^2}
\end{equation}
where $\rho$ is the effective signal to noise ratio in the presence
of correlated Gaussian noise with length scale $\theta_N$ and is given by
\begin{equation}
\rho^2 = \frac{\theta_M\theta_m}{4\theta^2_N}\left[
  1 + \left(\frac{\theta_N}{\theta_M} \right)^2 \right]^{\alpha_M} \left[ 
  1 + \left(\frac{\theta_N}{\theta_m} \right)^2 \right]^{\alpha_m} 
  \frac{A^2_{843}}{\sigma^2} 
\end{equation}
where $\theta_M$ is the fitted major axis and $\theta_m$ is the fitted minor
axis. $A_{843}$ is the peak brightness of the fitted Gaussian, and $\sigma$ is
the local rms noise as defined in Section \ref{s_noise}.

The relative flux density calibration was calculated by \citet{mauch03} 
using data from an analysis of the Molonglo calibrators by \citet{gaensler00}. 
The average scatter $\epsilon_{A_{843}}$ in repeated observations of the 
calibrators over a 12 year period was found to be around 3 per cent. 
This value was adopted as the relative calibration uncertainty of MOST peak
brightness measurements. The peak brightness uncertainties in the MGPS-2 
catalogue were obtained by adding $\epsilon_{A_{843}}\sigma_{A_{843}}$ and 
$A_{843}$ in quadrature. 

As an additional check on the likely variability of flux density measurements
for non-calibrator sources, we have looked at sources in one field of 
MGPS-2 that was repeatedly observed over the duration of the survey. 
This field, centred at 
$(\alpha, \delta) = 16^h58^m, -49\degr00\arcmin$ was 
observed 19 times between 2004 and 2007.
There were 28 sources with peak flux density $> 50$ mJy beam$^{-1}$
that appeared in each repeat observation of the field. Using these sources, 
we found the average rms difference to be around 5 per cent. This is 
slightly higher than for the Molonglo calibrator sources.

The integrated flux density of the source is calculated from the Gaussian 
fit performed by {\sc vsad}. We use the same method for deriving the 
integrated flux density as used for the NVSS \citep{condon98}. It should be 
noted that the quoted flux density uncertainties do not take into account the 
errors which arise from fitting extended or complex structures with 
an elliptical Gaussian model. Although significantly extended sources are 
not included in the MGPS-2 catalogue, this effect also applies to slightly extended 
sources, for which the quoted errors may not be reliable.

\section{Catalogue Format}\label{s_format}
Table 4 
shows the first page of the MGPS-2 catalogue.
The format is similar to the SUMSS catalogue format, with additional columns giving 
Galactic coordinates. A short description of the columns of the 
catalogue follows. 

\medskip

\noindent  \textit{Column} (1): The MGPS-2 source name.

\noindent  \textit{Columns} (2) \& (3): The Galactic longitude ($l$) and 
  latitude ($b$) of the source.

\noindent  \textit{Columns} (4) \& (5): The Right Ascension ($\alpha$) and 
  Declination ($\delta$) of the source in J2000 coordinates.
  
\noindent  \textit{Column} (6): The uncertainty in Right Ascension ($\Delta\alpha$) in 
  arcseconds, calculated from the quadratic sum of the MOST Right Ascension 
  calibration uncertainty  and equation~\ref{raerr}.
  
\noindent  \textit{Column} (7): The uncertainty in Declination ($\Delta\delta$) in
  arcseconds, calculated from the quadratic sum of the MOST Declination 
  calibration uncertainty  and equation~\ref{decerr}.

\noindent  \textit{Column} (8) \& (9): The peak brightness ($A_{843}$)
  in units of mJy beam$^{-1}$ and its associated uncertainty 
  ($\sigma_A$) calculated as described in Section \ref{s_flux}.
 
\noindent  \textit{Column} (10) \& (11): The total flux density ($S_{843}$)
  in units of mJy and its associated uncertainty ($\sigma_S$) calculated
  from the equations described in \cite{condon97}. 
  
\noindent  \textit{Columns} (12) \& (13): The fitted major \& minor axes 
  ($\theta_M$, $\theta_m$) of the source in arcseconds.
 
\noindent  \textit{Column} (14): The fitted major axis position angle (PA$_F$)
  of the source in degrees east of north. Most unresolved sources would have 
  PA$_F$ values close to 0$^{\circ}$ or 180$^{\circ}$ since the MOST elliptical 
  beam has PA$=0$.

\noindent  \textit{Column} (15): The major axis size ($\phi_M$) in arcseconds, 
  after deconvolution from the MOST beam --- given if the fitted major axis size 
  exceeds the beam size by more than $2.33\sigma(\theta_{\rm{M}})$.

\noindent   \textit{Column} (16): The deconvolved minor axis size ($\phi_m$) in
  arcseconds. If the major axis is resolved the minor axis is
  subsequently checked using the same criterion, and the deconvolved size given.

\noindent   \textit{Column} (17): Deconvolved position angle (PA$_S$) in degrees 
  east from north --- given if the major axis is resolved.
  
\noindent   \textit{Column} (18): The name of the mosaic in which the source 
  appears. The original mosaics are available online. In the case where a source 
  appears in two mosaics (due to overlap), the mosaic name quoted is the one for
  which the fitted parameters are listed in the catalogue.
  
\noindent   \textit{Column} (19): The number of mosaics in which the source 
  appears. Sometimes a source can appear in multiple mosaics because of a small
  overlap region used to ensure all sources are properly imaged and not truncated.
  The source parameters which appear in the catalogue are those for the most 
  reliable fit.
  
\noindent   \textit{Columns} (20) \& (21): The x \& y pixel positions of the 
  source on the quoted mosaic.

\medskip

MGPS-2 catalogue sources should be referred to by their full IAU
designation \citep{lortet94}. These are of the form MGPS {\it JHHMMSS$-$DDMMSS} where
MGPS is the survey acronym, {\it J} specifies J2000.0 coordinate equinox,
{\it HHMMSS} are the hours, minutes and truncated seconds of Right
Ascension, $-$ is the sign of Declination and {\it DDMMSS} are the degrees,
minutes and truncated seconds of Declination. These are given in column 1
of Table 4.

\section{Analysis}\label{s_analysis}

\subsection{Completeness}
The completeness of the survey in the $|b|>2\degr$ region should be the same
as for SUMSS, as identical techniques were used to produce both catalogues,
and that region is largely unaffected by Galactic plane emission. 
\cite{mauch03} found SUMSS to be better than 99 per cent complete above 
8 mJy beam$^{-1}$ in southern mosaics ($\delta \le -50\degr$) and 
above 18 mJy beam$^{-1}$ for more northern mosaics ($\delta > -50\degr$).

In the $|b|\le 2\degr$ region completeness is difficult to calculate
because of the presence of extended diffuse sources and the negative bowls
around large bright sources.
In regions of both high positive and substantial negative flux density
it would be impossible to detect point sources at the level of our 
cutoff (10 mJy beam$^{-1}$).

As a result we have taken a dual approach. We have first created
a mask for all regions in which it would be impossible to detect sources down
to our flux density limit. 
Inside the masked areas it is not possible to measure completeness. 
Compact sources detected in these regions may still be included in the 
catalogue, but users should note that the catalogue is incomplete in 
these regions.

Outside the masked areas we calculated completeness using a simulation.
We injected 1000 artificial sources into a selection of residual 
mosaics (mosaics after all sources fitted by {\sc vsad} had been removed). 
The flux distribution of the sources followed that given in 
\citet{large91}. We then calculated the fraction of these artificial sources 
that were found by our cataloguing process. At the flux density cutoff 
(10 mJy beam$^{-1}$) the completeness is approximately 80 per cent, 
rising to greater than 99 per cent at 20 mJy beam$^{-1}$.

\subsection{MGPS-MRC cross-match}
As an additional check of catalogue completeness for brighter sources, we have 
crossmatched the MGPS-2 catalogue with the Molonglo Reference Catalogue (MRC)
\citep{large81,large91}. The MRC covers the southern sky, 
excluding the Galactic plane ($|b| < 3\degr$). The MRC was made from
$2\farcm6\times2\farcm9\,\rm{sec}\left(\delta+35\fdg5\right)$ resolution 
observations at 408 MHz using the Molonglo Cross radio telescope, 
the previous incarnation of the MOST. 
It is complete to $S_{408}=1$~Jy at 408~MHz and has a
limiting flux of $S_{408}=0.7$~Jy.
Assuming a typical spectral index of $-0.8$ we
expect the faintest MRC sources to appear in the MGPS-2 catalogue at around
$S_{843}=400$~mJy, which implies that all MRC sources within the MGPS-2 survey
region should appear in our catalogue. 

There are a total of 456 MRC source that fall in the region of the MGPS-2.
Of these, 453 sources had a MGPS-2 match within $1\arcmin$. Of the 3 sources 
with no match, two were excluded from our catalogue as they were extended 
sources:
\begin{itemize}
\item MRC B1610$-$605, a head-tail radio galaxy \citep{jones94}
\item MRC B0821$-$428, the Puppis-A supernova remnant \citep{green71}
\end{itemize}
The other object (MRC B0819$-$300B) was resolved as a double source by the MOST 
beam and so has two counterparts in MGPS-2. 
Hence the MGPS-2 catalogue is complete for bright sources.

\subsection{Compact source density}\label{s_density}
Analysis from earlier radio surveys of the Galactic plane suggested
the presence of a small overdensity of compact sources along the Galactic plane 
\cite[e.g.][]{whiteoak92a}. In this section we investigate the variation in 
density of the the MGPS-2 compact sources with Galactic latitude.

Calculating the source density in the Galactic plane region is non-trivial
because of the large number of diffuse extended objects, and the regions of 
negative emission around strong sources. 
We started with the assumption that for all regions covered with emission 
$ \ge 10$ mJy beam$^{-1}$ and all negative bowls 
$ < 0$ mJy beam$^{-1}$ it would be impossible to detect sources 
at our 10 mJy beam$^{-1}$ cutoff level.
Using the flood fill method described in Section \ref{s_mgpscat} we were able 
to identify all extended regions of positive or negative emission, which were then
excluded when calculating the sky area.

In regions away from the plane (i.e. $2\degr < |b| < 10\degr$) we would 
expect the compact source density to agree with the predicted extragalactic 
source density. \citet{mauch03} found a source density of 
$21.1\pm0.1\,\rm{deg}^{-2}$ over a $\sim1680\,\rm{deg}^2$ area of sky which 
the SUMSS survey had completed at the time of publication.
An updated value that we have calculated using the entire area of the SUMSS
survey, is $21.87\pm0.05\,\rm{deg}^{-2}$.

\begin{figure}
\begin{center}
\includegraphics[width=\linewidth]{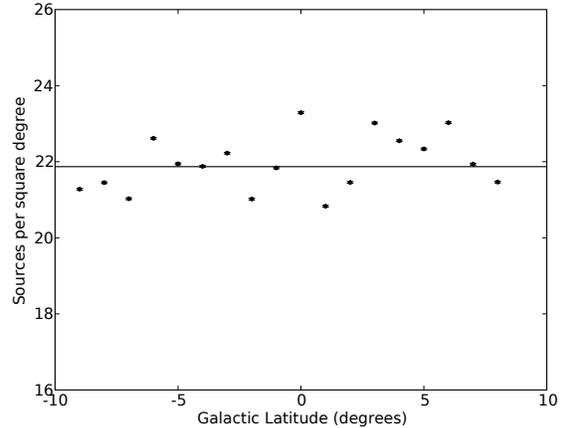}
\caption{Compact source density as a function of Galactic latitude. 
The error in number of sources $N$ in each latitude bin is plotted as the 
Poisson error $\frac{1}{\sqrt{N}}$ --- the error bars are approximately the 
same size as the point markers in the plot.
The standard deviation of the points is $\sigma = 1.3$. The horizontal 
line at 21.87 shows the extragalactic source density calculated using SUMSS.
\label{f_density}}
\end{center}
\end{figure}

\changes{
The $\pm\,0.1$ error quoted for the SUMSS source densities is based on 
Poisson statistics alone. A more accurate estimate of the error (which 
includes systematic errors, effects due to clustering and random fluctuations 
in the source density) is given by the variance between measurements of the 
source density in different regions of the sky.
Figure \ref{f_density} shows the source density for point-like sources (i.e. 
those included in this catalogue) as a function of Galactic latitude. 
The scatter in the data points clearly exceeds that predicted by Poisson 
statistics --- $\frac{1}{\sqrt{N}}$ where $N$ is the number of sources in 
the latitude bin. The standard deviation of the data points is $\sigma=1.3$, 
which gives a more suitable estimate of the variance than the Poisson error alone. 
For the region $2\degr < |b| < 10\degr$ we calculate a source density of 
$21.8\pm1.3\,\rm{deg}^{-2}$. 
Considering this scatter, our estimate of the extragalactic source density is 
in agreement with that from \citet{mauch03} and our latest calculations using the 
full SUMSS catalogue.
}

As shown in Figure \ref{f_density}, we find no significant overdensity of compact sources
in the Galactic plane. This is in agreement with \citet{helfand06} who find 
no significant dependence on Galactic latitude for their compact source 
counts. Inspection of Galactic plane observations (such as in Figures 
\ref{f_strips1} and \ref{f_strips2}) clearly reveal an overdensity of sources 
close to the Galactic equator. In light of this, our results are not 
intuitive, but they can be explained.
Firstly, small diameter Galactic sources (e.g. young supernova remnants and 
ultra-compact HII regions) are not distributed evenly throughout the Galaxy.
They tend to occur in and around large complexes of star formation 
\citep{weiler88}.
Many of these regions of extended diffuse emission are masked out and 
excluded from our analysis, as described above.
Secondly, the expected number density of most small diameter 
Galactic sources is less than the systematic errors in our source density values. 
For example, in the ATNF Pulsar 
Catalogue\footnote{{\tt http://www.atnf.csiro.au/research/pulsar/psrcat/}} of 
\citet{manchester05} there are 635 pulsars that fall within the bounds of MGPS-2. 
\changes{
Of these, we have detected 93 that have $A_{843} \ge 10$ mJy beam$^{-1}$ 
and that fall in a non-masked region. 
This is a number density of $\sim 0.05\,\rm{deg}^2$, 
which is well below the systematic error levels in our results above.
Likewise for supernova remnants, \citet{helfand89} predict (as a rough 
estimate) that there are around 200 missing supernovae in the Galactic plane. 
This is an upper limit for our purposes, since many of these are likely 
to be too extended to include in our compact source catalogue. So, again, 
this number of additional Galactic sources would fit well below the 
systematic error levels of our source density calculation.
}

\section{Summary and Future Work}
The main data product of MGPS-2 are 130 $4\fdg3 \times 4\fdg3$ mosaics which we
have released online.
Note that some of the mosaics fall across the $|b| = 10\degr$ survey
boundary and have been released as part of the SUMSS survey.

We have created a catalogue of $48\,850$ compact sources covering the
2400\,deg$^2$ of the MGPS-2 survey.
A future release will include diffuse and extended sources.
The limiting peak brightness of the catalogue is 10 mJy beam$^{-1}$.
We have calculated the compact source density in bins of Galactic latitude and
found it agrees within the errors with the expected source density for 
extragalactic sources. This implies that even close to the Galactic plane 
the majority of compact sources in the radio sky are extragalactic.
The reason we see no overdensity is that highly complex regions of emission
were masked out and excluded from the analysis, and this is where compact HII 
regions and young supernova remnants are likely to occur.

Users of the catalogue should note that for some sources (especially those 
which are slightly resolved) the quoted uncertainties may be underestimates
of the true uncertainties.
Users are encouraged to check the mosaics directly if in doubt.

Following on from this compact source catalogue there will be a catalogue and 
census of extended sources in the Galactic plane, particularly 
young supernova remnants and pulsar wind nebulae.
Future work with MGPS-2 will look at the physical processes producing
the observed large scale structures, and look at structures in which the 
radio--FIR correlation breaks down. We will use the current catalogue to search
for ultra-compact HII regions, believed to be signposts of the early stages
of massive star formation.

The MOST itself is undergoing redevelopment to produce a new low frequency
spectral line capability with dual polarisation feeds. As the Square Kilometre
Array Molonglo Prototype \citep[SKAMP;][]{adams04}, it will be used to map
out magnetic field structures and test theories of mass assembly of galaxies
at $z\sim0.7$ when the Universe was roughly half its current age.

\section*{Acknowledgments}
The Molonglo Observatory site manager, Duncan Campbell-Wilson, and the staff 
Rodger Ashwell, Adrian Blake and Tim Hubbard, and former staff Jeff Webb, Michael 
White and John Barry, are responsible for the smooth operation of the MOST 
telescope and the day to day observing program of the SUMSS and MGPS-2 surveys. 
We would also like to thank Tony Turtle for his work on the scheduling of the
surveys.
The MOST is operated with the support of the Australian Research 
Council and the School of Physics within the University of Sydney.
Tara Murphy acknowledges the support of an ARC Australian Postdoctoral Fellowship.

\bibliographystyle{mn2e}
\bibliography{mn-jour,mgps-fix}

\include{landtable}

\begin{twocolumn}

\begin{figure*}
\begin{center}
\includegraphics[angle=90,width=16cm]{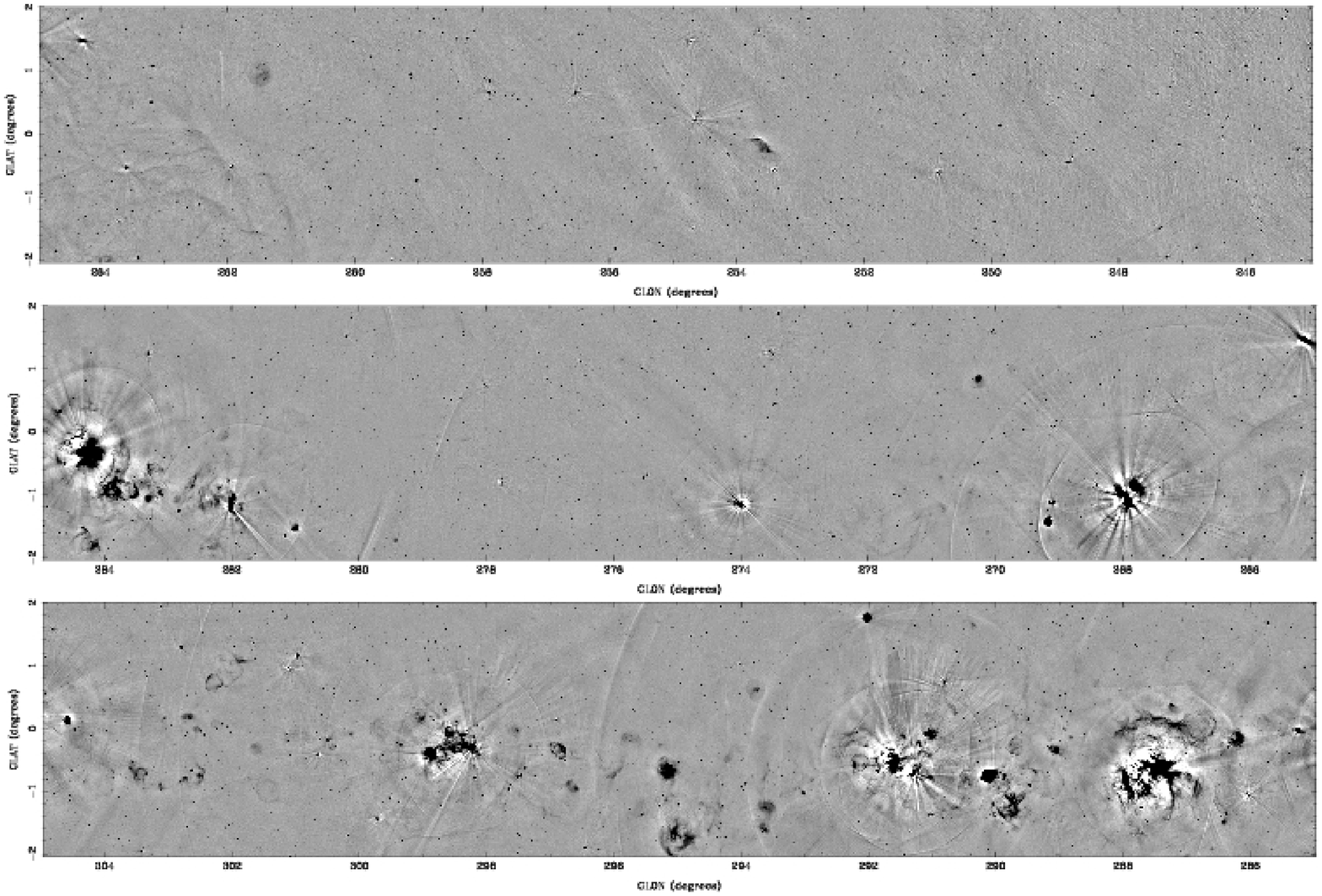}
\caption{Panoramic images of the Galactic plane from MGPS-2 covering 
$|b| < 2\degr$, $245\degr \le l \le 305\degr$. The greyscale ranges from 
$-26$ to 45 mJy beam$^{-1}$.\label{f_strips1}}
\end{center}
\end{figure*}

\begin{figure*}
\begin{center}
\includegraphics[angle=90,width=16cm]{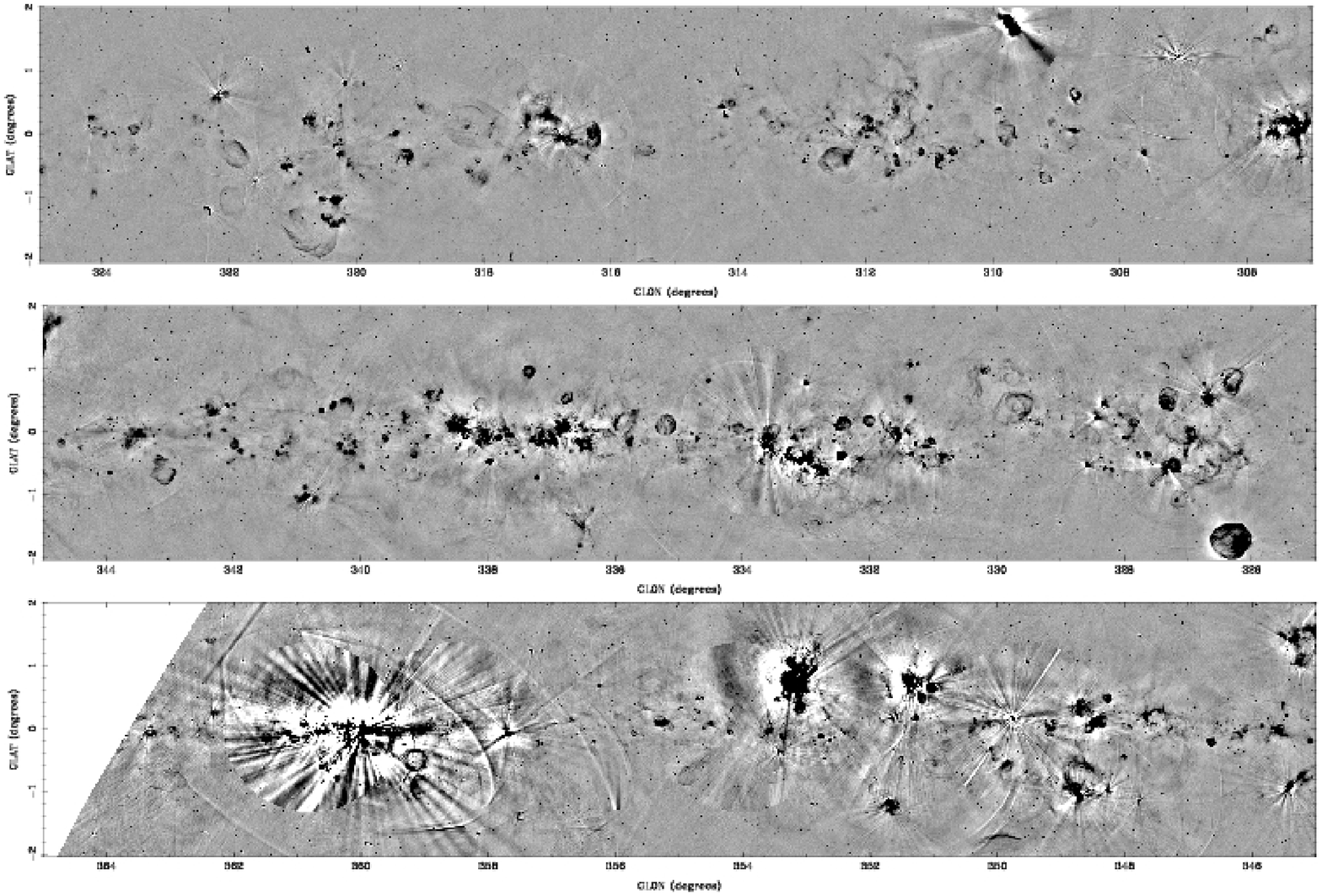}
\caption{Panoramic images of the Galactic plane from MGPS-2, covering
$|b| < 2\degr$, $305\degr \le l \le 365\degr$. The greyscale ranges from 
$-26$ to 45 mJy beam$^{-1}$.\label{f_strips2}}
\end{center}
\end{figure*}

\end{twocolumn}

\label{lastpage}
\end{document}

%% file: landtable.tex
\setcounter{table}{4}
\begin{onecolumn}
\begin{sidewaystable}
\begin{minipage}{225mm}
{\bf Table 4.} The first page of the MGPS-2 compact source catalogue. \\
\label{t_cat}
\setlength{\tabcolsep}{1mm}
\begin{tabular}{@{}ccccrrrrrrrrrrrrcccrr}
\hline
(1) & (2) & (3) & (4) & (5) & (6) & (7) & \multicolumn{1}{c}{(8)} & \multicolumn{1}{c}{(9)} & \multicolumn{1}{c}{(10)} & \multicolumn{1}{c}{(11)} & \multicolumn{1}{c}{(12)} & \multicolumn{1}{c}{(13)} & \multicolumn{1}{c}{(14)} & \multicolumn{1}{c}{(15)} &
\multicolumn{1}{c}{(16)} & (17) & \multicolumn{1}{c}{(18)} & (19) & (20) & (21)  \\

IAU designation & $l$ & $b$ &
\(\alpha\) (J2000) & \(\delta\) (J2000) & \(\Delta\alpha\) & \(\Delta\delta\) & \multicolumn{1}{c}{\(A{_{843}^{\rm{a}}}\)} &
\(\sigma{_A}\) & \multicolumn{1}{c}{\(S{_{843}^{\rm{b}}}\)} & \(\sigma{_S}\) &
\multicolumn{1}{c}{\(\theta{_{\rm{M}}}^{\rm{c}}\)} &
\multicolumn{1}{c}{\(\theta{_{\rm{m}}}^{\rm{c}}\)} &
\multicolumn{1}{c}{\(\rm{PA}{_{\rm{F}}}^{\rm{c,d}}\)} & 
\multicolumn{1}{c}{\(\phi{_{\rm{M}}}^{\rm{e}}\)} & \multicolumn{1}{c}{\(\phi{_{\rm{m}}}^{\rm{e}}\)}
& $\rm{PA}{_{\rm{S}}}^{\rm{e,d}}$ & Mosaic\(^{\rm{f}}\) & \#\(^{\rm{f}}\) & x-pix & y-pix \\

& $\degr$ & $\degr$ & \(h \,\,\, m \,\,\, s\) & \(^{\circ} \,\,\, \arcmin \,\,\, \arcsec\) & \arcsec & \arcsec &
\multicolumn{2}{c}{mJy beam\(^{-1}\)} & \multicolumn{2}{c}{mJy} &
\multicolumn{1}{c}{\arcsec} & \multicolumn{1}{c}{\arcsec}
& \multicolumn{1}{c}{\(^\circ\)} & \multicolumn{1}{c}{\arcsec} &
\multicolumn{1}{c}{\arcsec} & \multicolumn{1}{c}{\(^\circ\)} & & & & \\
\hline
MGPS J071450$-$330700 & 245.155 & $-$9.972 & 07 14 50.01 & $-$33 07 00.6 & 2.2 & 2.4 & 24.6 & 1.3 & 24.6 & 1.3 & 84.9 & 58.5 & 138.5 & 0.0 & 0.0 & --- & J0712M32 & 1 & 510.8 & 180.0 \\
MGPS J071512$-$331111 & 245.254 & $-$9.932 & 07 15 12.51 & $-$33 11 11.4 & 2.3 & 2.1 & 22.7 & 1.3 & 22.7 & 1.3 & 84.9 & 49.2 & 118.1 & 0.0 & 0.0 & --- & J0712M32 & 1 & 485.3 & 167.8 \\
MGPS J071525$-$330343 & 245.161 & $-$9.836 & 07 15 25.71 & $-$33 03 43.3 & 2.3 & 2.7 & 22.5 & 1.3 & 22.5 & 1.3 & 84.9 & 63.4 & 156.9 & 0.0 & 0.0 & --- & J0712M32 & 1 & 469.9 & 189.3 \\
MGPS J071536$-$332226 & 245.461 & $-$9.940 & 07 15 36.16 & $-$33 22 26.5 & 2.1 & 2.3 & 28.0 & 1.3 & 31.4 & 1.5 & 84.9 & 66.1 & 139.3 & 0.0 & 0.0 & --- & J0712M32 & 1 & 458.9 & 135.2 \\
MGPS J071600$-$325703 & 245.113 & $-$9.679 & 07 16 00.00 & $-$32 57 03.5 & 2.3 & 2.5 & 28.3 & 1.4 & 31.4 & 1.6 & 84.9 & 68.7 & 143.8 & 0.0 & 0.0 & --- & J0712M32 & 1 & 430.5 & 208.4 \\
MGPS J071601$-$325124 & 245.029 & $-$9.633 & 07 16 01.48 & $-$32 51 24.5 & 1.5 & 1.8 & 102.1 & 3.3 & 120.9 & 3.9 & 90.6 & 55.0 & 167.7 & 0.0 & 0.0 & --- & J0712M32 & 1 & 428.4 & 224.7 \\
MGPS J071650$-$340139 & 246.172 & $-$9.997 & 07 16 50.22 & $-$34 01 39.0 & 3.5 & 2.5 & 10.5 & 1.0 & 10.5 & 1.0 & 84.9 & 46.0 & 75.2 & 0.0 & 0.0 & --- & J0712M32 & 2 & 377.0 & 21.5 \\
MGPS J071657$-$330639 & 245.348 & $-$9.571 & 07 16 57.16 & $-$33 06 39.3 & 1.5 & 1.8 & 67.0 & 2.2 & 67.0 & 2.2 & 84.9 & 47.5 & 5.8 & 0.0 & 0.0 & --- & J0712M32 & 1 & 365.6 & 180.3 \\
MGPS J071705$-$335242 & 246.059 & $-$9.885 & 07 17 05.01 & $-$33 52 42.0 & 2.2 & 2.6 & 23.1 & 1.2 & 23.1 & 1.2 & 84.9 & 62.2 & 20.7 & 0.0 & 0.0 & --- & J0712M32 & 1 & 359.7 & 47.2 \\
MGPS J071707$-$325429 & 245.179 & $-$9.450 & 07 17 07.05 & $-$32 54 29.2 & 3.9 & 3.0 & 11.2 & 1.1 & 11.8 & 1.2 & 84.9 & 49.8 & 69.7 & 0.0 & 0.0 & --- & J0712M32 & 1 & 353.5 & 215.4 \\
MGPS J071730$-$334543 & 245.992 & $-$9.755 & 07 17 30.22 & $-$33 45 43.3 & 3.5 & 3.6 & 16.6 & 1.2 & 24.0 & 1.8 & 84.9 & 76.2 & 127.3 & 0.0 & 0.0 & --- & J0712M32 & 1 & 330.7 & 67.2 \\
MGPS J071731$-$324109 & 245.015 & $-$9.275 & 07 17 31.47 & $-$32 41 09.1 & 4.1 & 4.1 & 17.6 & 1.7 & 21.4 & 2.1 & 84.9 & 65.8 & 130.1 & 0.0 & 0.0 & --- & J0712M32 & 1 & 324.6 & 253.8 \\
MGPS J071755$-$331653 & 245.595 & $-$9.463 & 07 17 55.82 & $-$33 16 53.1 & 2.1 & 3.2 & 13.3 & 1.0 & 13.3 & 1.0 & 84.9 & 46.5 & 174.2 & 0.0 & 0.0 & --- & J0712M32 & 1 & 299.4 & 150.3 \\
MGPS J071811$-$334824 & 246.096 & $-$9.648 & 07 18 11.14 & $-$33 48 24.9 & 2.8 & 3.2 & 21.9 & 1.5 & 21.9 & 1.5 & 84.9 & 67.2 & 152.2 & 0.0 & 0.0 & --- & J0712M32 & 1 & 284.5 & 59.1 \\
MGPS J071819$-$334633 & 246.081 & $-$9.609 & 07 18 19.33 & $-$33 46 33.4 & 1.5 & 1.7 & 367.5 & 11.1 & 384.2 & 11.6 & 84.9 & 59.0 & 134.5 & 0.0 & 0.0 & --- & J0712M32 & 1 & 275.1 & 64.4 \\
MGPS J071903$-$332957 & 245.899 & $-$9.349 & 07 19 03.47 & $-$33 29 57.7 & 4.8 & 4.5 & 10.7 & 1.3 & 11.3 & 1.4 & 84.9 & 55.5 & 51.2 & 0.0 & 0.0 & --- & J0712M32 & 1 & 223.5 & 111.9 \\
MGPS J071904$-$345732 & 247.227 & $-$9.994 & 07 19 04.18 & $-$34 57 32.2 & 3.4 & 4.2 & 13.9 & 1.4 & 13.9 & 1.4 & 76.6 & 56.6 & 20.3 & 0.0 & 0.0 & --- & J0712M36 & 1 & 231.0 & 612.7 \\
MGPS J071906$-$342641 & 246.762 & $-$9.759 & 07 19 06.50 & $-$34 26 41.4 & 2.6 & 4.0 & 14.5 & 1.4 & 17.2 & 1.7 & 88.8 & 48.1 & 169.6 & 0.0 & 0.0 & --- & J0712M36 & 1 & 225.5 & 711.6 \\
MGPS J071912$-$334933 & 246.209 & $-$9.466 & 07 19 12.43 & $-$33 49 33.6 & 4.9 & 4.3 & 10.5 & 1.4 & 10.5 & 1.4 & 84.9 & 51.5 & 124.0 & 0.0 & 0.0 & --- & J0712M32 & 1 & 215.2 & 55.2 \\
MGPS J071918$-$340657 & 246.481 & $-$9.577 & 07 19 18.23 & $-$34 06 57.4 & 1.7 & 1.8 & 71.3 & 2.7 & 77.8 & 2.9 & 76.6 & 50.5 & 105.8 & 0.0 & 0.0 & --- & J0712M36 & 1 & 210.3 & 774.6 \\
MGPS J071922$-$341214 & 246.568 & $-$9.603 & 07 19 22.74 & $-$34 12 14.2 & 1.5 & 1.7 & 172.7 & 5.5 & 186.8 & 5.9 & 76.6 & 58.2 & 131.6 & 0.0 & 0.0 & --- & J0712M36 & 1 & 205.8 & 757.7 \\
MGPS J071925$-$340056 & 246.402 & $-$9.509 & 07 19 25.81 & $-$34 00 56.2 & 1.5 & 1.7 & 186.0 & 5.8 & 199.1 & 6.2 & 84.9 & 50.3 & 108.6 & 0.0 & 0.0 & --- & J0712M32 & 2 & 201.2 & 22.2 \\
MGPS J071944$-$322754 & 245.027 & $-$8.758 & 07 19 44.61 & $-$32 27 54.2 & 7.5 & 5.9 & 15.0 & 2.7 & 18.2 & 3.3 & 84.9 & 58.1 & 112.7 & 0.0 & 0.0 & --- & J0712M32 & 1 & 170.6 & 290.8 \\
MGPS J071954$-$341207 & 246.616 & $-$9.504 & 07 19 54.29 & $-$34 12 07.2 & 1.5 & 1.7 & 302.3 & 9.3 & 331.6 & 10.2 & 76.6 & 57.5 & 128.1 & 0.0 & 0.0 & --- & J0712M36 & 2 & 170.2 & 757.6 \\
MGPS J071959$-$350449 & 247.422 & $-$9.878 & 07 19 59.57 & $-$35 04 49.3 & 2.2 & 1.9 & 37.7 & 2.0 & 37.7 & 2.0 & 76.6 & 45.0 & 87.1 & 0.0 & 0.0 & --- & J0712M36 & 2 & 169.9 & 588.7 \\
MGPS J072002$-$342120 & 246.768 & $-$9.548 & 07 20 02.29 & $-$34 21 20.6 & 5.3 & 4.3 & 12.3 & 1.8 & 13.0 & 1.9 & 76.6 & 50.5 & 64.4 & 0.0 & 0.0 & --- & J0712M36 & 2 & 162.2 & 728.0 \\
MGPS J072004$-$343735 & 247.018 & $-$9.660 & 07 20 04.90 & $-$34 37 35.5 & 4.4 & 3.0 & 11.1 & 1.4 & 11.1 & 1.4 & 76.6 & 45.1 & 92.4 & 0.0 & 0.0 & --- & J0712M36 & 2 & 161.0 & 675.9 \\
MGPS J072005$-$351416 & 247.574 & $-$9.930 & 07 20 05.36 & $-$35 14 16.0 & 6.5 & 5.8 & 10.1 & 1.9 & 10.1 & 1.9 & 76.6 & 51.1 & 124.8 & 0.0 & 0.0 & --- & J0712M36 & 1 & 164.5 & 558.3 \\
MGPS J072006$-$343027 & 246.912 & $-$9.602 & 07 20 06.75 & $-$34 30 27.0 & 3.1 & 3.1 & 19.9 & 1.6 & 20.9 & 1.7 & 76.6 & 55.7 & 129.5 & 0.0 & 0.0 & --- & J0712M36 & 2 & 158.2 & 698.7 \\
MGPS J072009$-$335332 & 246.358 & $-$9.320 & 07 20 09.20 & $-$33 53 32.3 & 2.2 & 2.2 & 46.7 & 2.4 & 51.4 & 2.7 & 84.9 & 57.1 & 124.4 & 0.0 & 0.0 & --- & J0712M32 & 4 & 151.4 & 43.1 \\
\hline
\end{tabular}

\flushleft
NOTES:\\
\(^{\rm{a}}\) The peak brightness of the Gaussian fit in units of mJy beam\(^{-1}\). This value may be
in error by more than the quoted error for extended sources.\\
\(^{\rm{b}}\) The total flux density of the Gaussian fit in units of
mJy. \(S = A\) for point sources.\\
\(^{\rm{c}}\) The widths and position angle of the fitted Gaussian. The fit
is constrained so that \(\theta_{\rm{m}}\geq45\arcsec\) (the beam minor axis width).\\
\(^{\rm{d}}\) The position angle of the major axis is measured in degrees east from 
north.\\
\(^{\rm{e}}\) The deconvolved widths and position angle of the source. A
value is given only if the fitted axis exceeds the beam by more than 2.33\(\sigma_{\theta}\)\\
\(^{\rm{f}}\) The name of the mosaic the quoted source can be found
in. If the number in the next column is greater than 1 it can also be found in neighbouring 
mosaics.\\

\end{minipage}

\end{sidewaystable}

\end{onecolumn}
